%% file: arxiv.tex
\documentclass[12pt,a4paper,reqno]{article}

\usepackage{authblk}
\usepackage{geometry}
\usepackage[authoryear,round]{natbib}

\include{header}

\author[1,4]{S. Jeyakumar}
\author[2,3]{M. Kraus}
\author[1,5]{M. J. Hole}
\author[4]{D. Pfefferlé}

\affil[1]{\small Mathematical Sciences Institute, Australian National University, Acton ACT 2601, Australia}
\affil[2]{\small Max Planck Institute for Plasma Physics, Boltzmannstra{\ss}e 2, 85748 Garching, Germany}
\affil[3]{\small Technical University of Munich, Department of Mathematics, Boltzmannstra{\ss}e 3, 85748 Garching, Germany}
\affil[4]{\small The University of Western Australia, 35 Stirling Highway, Crawley WA 6009, Australia}
\affil[5]{\small Australian Nuclear Science and Technology Organisation, Locked Bag 2001, Kirrawee DC, NSW 2232, Australia}

\begin{document}

\maketitle

\input{abstract}

\tableofcontents

\pagebreak

\input{body}


\bibliographystyle{jpp}

\bibliography{refs} 

\end{document}

%% file: header.tex
\usepackage{graphicx}

\usepackage[utf8]{inputenc}
\usepackage[T1]{fontenc}
\usepackage{amsmath}
\usepackage{amssymb}
\usepackage{amsthm}

\theoremstyle{definition}

\newcommand{\R}{\mathbb{R}}

\newcommand{\M}{\mathbb{M}}
\newcommand{\G}{\mathbb{G}}
\newcommand{\dx}[1]{\text{d}#1}

\title{Structure-preserving particle methods for the Landau collision operator using the metriplectic framework}

%% file: abstract.tex
\begin{abstract}

We present a novel family of particle discretisation methods for the nonlinear Landau collision operator.
We exploit the metriplectic structure underlying the Vlasov--Maxwell--Landau system in order to obtain disretisation schemes that automatically preserve mass, momentum, and energy, warrant monotonic dissipation of entropy, and are thus guaranteed to respect the laws of thermodynamics.
In contrast to recent works that used radial basis functions and similar methods for regularisation, here we use an auxiliary spline or finite element representation of the distribution function to this end.
Discrete gradient methods are employed to guarantee the aforementioned properties in the time discrete domain as well.

\end{abstract}

%% file: body.tex
\section{Introduction}

In this work, we present a structure-preserving particle based approach to simulating the Landau collision operator, using the metriplectic formulation of \cite{Morrison1986}. The subject of structure-preserving algorithms to simulate ideal dynamics in plasmas has been well studied \citep{Morrison2017}. In the context of ideal kinetics in particular, such as the Vlasov--Maxwell or Vlasov--Poisson systems, recent works include \citet{Chen2011, Markidis2011, Squire2012, Evstatiev:2013, Qin2016, Burby2017, Kraus2017, Zhang2017, CamposPinto2022}.
Complementing the ideal dynamics with dissipative effects, e.g. those caused by collisions, complicates the construction of structure-preserving numerical methods for the coupled system and is a more recent topic of study. 

Initial works studying dissipation were primarily based on grid-based approaches \citep{Yoon2014, Taitano2015, Hirvijoki2017a, Kraus2017b, Shiroto2019}, however, there have been some recent advances in studying particle-based structure-preserving methods for this application. \citet{hirvijoki2018metriplectic} approach this by varying the weights of the marker particles, instead of their velocities. \citet{Carrillo2020} and \citet{Hirvijoki2021} use finite-sized marker particles to discretise the Landau operator, in particular by convolving the delta-function particles with a finite-sized shape function. Further work by \cite{Bailo2024} extends such a regularisation approach to the full Vlasov--Maxwell--Landau system. 

Alternative approaches include Monte-Carlo methods, which treat collisions as a stochastic process and effectively model their underlying microscopic behaviour (see for e.g. \cite{Boozer1981}). More recent versions of such an approach are structure-preserving stochastic methods for collision operators, including the Landau operator \citep{Tyranowski2021a, Tyranowski2021}. \cite{Schnake2024} considers a sparse-grid discontinuous Galerkin approach to the Vlasov--Poisson--Lenard--Bernstein equations, which considers both a sparse grid for the entire phase space, as well as a hybrid grid which only uses a sparse grid in velocity space. Finally, the work of \cite{Laidin2024} considers a constrained polynomial based approach to solving collisional equations while conserving the required moments, such as momentum and energy. 

This manuscript is structured as follows.
In Section~\ref{sec:metriplectic}, we give a short overview of the metriplectic framework, the Vlasov--Maxwell--Landau system and how the Landau collision operator can be described in terms of a metric bracket. We pay special attention to the conservation properties and show how systems that fit into the metriplectic framework are automatically guaranteed to satisfy the laws of thermodynamics.
In Section~\ref{sec:particle-discretisation}, we describe the discretisation of the Landau operator by exploiting its metriplectic structure. We show how an auxiliary spline or finite element representation of the distribution function can be used for regularisation, prove conservation of mass, momentum, and energy, and show that entropy is dissipated monotonically, thereby verifying adherence of the semi-discrete scheme to the laws of thermodynamics.
In Section~\ref{sec:time-discretisation}, we show how the discrete gradient method can be used to obtain a discretisation in time that retains all aforementioned conservation properties of the semi-discrete scheme.

\section{The Metriplectic formulation}
\label{sec:metriplectic}

The metriplectic framework \citep{Morrison1986, Kaufman:1982fl, Kaufman:1984fb, Morrison:1984ca, Morrison:1984wu, Grmela:1984dn, Grmela:1984ea, Grmela:1985jd, morrison2023inclusive} provides a convenient way to express dynamical systems which consist of two parts: a conservative, Hamiltonian part and a dissipative part, which are compatible with the laws of thermodynamics. The ideal dynamics are described via a Hamiltonian functional and a Poisson bracket, and the dissipative dynamics through a symmetric metric bracket and a functional representing the dissipated quantity, usually the entropy. 

In this section, we give an overview of the metriplectic framework, briefly sketch the Vlasov--Maxwell--Landau system, and then explain the Landau collision operator in more detail.
For a slightly more comprehensive exposition of the metriplectic framework in the context of collision operators see e.g. \citet[Section 2.1]{Kraus2017b}.

\subsection{General Framework}

Let us denote by $u (t,z) = (u^{1}, \, u^{2}, \, ..., \, u^{m})^{T}$ the field variables, defined over the domain $\Omega$ with coordinates $z$, and let $\mathcal A$ be an arbitrary functional of the field
variables $u$.
In the metriplectic framework, the evolution of any functional $\mathcal{A}$ of the dynamical variables is given by:
\begin{equation}\label{eq:metriplectic_evolution_1}
    \dot{\mathcal{A}} = \{\mathcal{A} , \mathcal F\} + (\mathcal{A}, \mathcal F) ,
\end{equation}
where $\mathcal F = \mathcal H - \mathcal S$ is the free energy, i.e., the difference between the Hamiltonian functional $\mathcal H$ and the entropy functional $\mathcal S$.
The Poisson bracket $\{ \cdot, \cdot \}$ describes the conservative part of the dynamics.
It is a bilinear, anti-symmetric operation that satisfies Leibniz' rule and the Jacobi identity.
The metric bracket $( \cdot, \cdot )$ describes the dissipative part of the dynamics.
It is a bilinear, positive semi-definite, symmetric operation that satisfies Leibniz' rule.
With the sign convention chosen here, entropy is always dissipated.

The two brackets have to satisfy certain compatibility conditions which ultimately guarantee compliance with the laws of thermodynamics. The entropy functional is usually chosen to be a Casimir invariant of the Poisson bracket, meaning that it satisfies $\{\mathcal S, \mathcal A\} = 0$ for all functionals $\mathcal A$. The metric bracket is constructed to ensure that the Hamiltonian and other physically relevant invariants $\mathcal I$ of the conservative system (Casimir invariants or momentum maps) are Casimir invariants of the metric bracket, i.e. $(\mathcal I, \mathcal A) = 0$ for all functionals $\mathcal A$. Thus, the evolution equation of \eqref{eq:metriplectic_evolution_1} becomes:
\begin{equation}
    \dot{\mathcal{A}} = \{\mathcal{A} , \mathcal H\} - (\mathcal{A}, \mathcal S). 
\end{equation}

The compatibility conditions ensure a separation of the conservative and dissipative dynamics, as the Hamiltonian dynamics do not affect the entropy, and the dissipative dynamics do not affect the Hamiltonian or any other physically relevant invariant of the ideal dynamics. It is desirable to also preserve this property in the course of discretisation. 

\subsection{H-Theorem and Energy-Casimir principle}

With the aforementioned requirements that (i) $\mathcal H$ is a Casimir of the metric bracket, (ii) $\mathcal S$ is a Casimir of the Poisson bracket, and that (iii) the metric bracket is positive semi-definite, the metriplectic framework automatically reproduces the First and Second Law of Thermodynamics:
\begin{subequations}
\begin{align}
\dfrac{d\mathcal H}{dt} &= \{\mathcal H,\mathcal H\} - (\mathcal H,\mathcal S) = 0 ,\\
\dfrac{d\mathcal S}{dt} &= \{\mathcal S,\mathcal H\} - (\mathcal S,\mathcal S) = - (\mathcal S,\mathcal S) \le 0 .
\end{align}
\end{subequations}

When a metriplectic system possesses no Casimir invariants, its equilibrium state, $u_{eq}$, is determined fully by an energy principle, $\delta \mathcal F[u_{eq}] = 0$. Additionally, this state is stable given the second variation satisfies $\delta^2 \mathcal F[u_{eq}] > 0$. When there are Casimir invariants, however, the equilibrium condition must be modified to account for the degeneracies in the state of the system~\citep{Morrison1998}. In this case, the equilibrium state must satisfy an energy-Casimir principle as follows:
\begin{equation}\label{eq:energy_casimir}
    \bigg[ \delta \mathcal F + \sum_i \lambda_i \delta \mathcal C_i \bigg] \bigg\vert_{u = u_{eq}} = 0, 
\end{equation}
where $\mathcal C_i$ are the Casimir invariants, and $\lambda_i$ are corresponding Lagrange multipliers which are determined by the initial conditions of the system. 

\subsection{The Vlasov-Maxwell-Landau system}

The Vlasov-Maxwell-Landau system consists of: the Vlasov equation, which describes the evolution of the phase-space probability distribution, Maxwell's equations, which describe the evolution of the electromagnetic fields, and the Landau operator, which describes collisional interactions between charged particles.
This system is described by the following set of equations: 
\begin{align}
    &\partial_t f + v\cdot \nabla_xf + \frac{q}{m}(E + v \times B) \cdot \nabla_vf = C[f], \\    
    &C[f] = \frac{\partial}{\partial v} \cdot \int U(v-  v') \left( f(v')\frac{\partial f}{\partial v} - f(v)\frac{\partial f}{\partial v'}\right) \dx  v', \\
    &\frac{1}{c^2} \frac{\partial E}{\partial t} = \text{curl }B - \mu_0q \int vf\text{d}v,\\
    &\frac{\partial B}{\partial t} = - \text{curl }E.
\end{align}
Here, $f(x,v,t) : \R^3 \times \R^3 \times [0, \infty) \to [0, \infty)$ is a non-negative scalar function representing the single-particle distribution function, and $(x,v) \in \R^3 \times \R^3$ are the position and velocity coordinates. The electric and magnetic fields are given by $E(x,t): \R^3 \times[0, \infty) \to \R^3 $ and $B(x,t): \R^3 \times[0, \infty) \to \R^3 $, respectively. The tensor $U(v - v')$ is a scaled orthogonal projector onto the orthogonal subspace of the vector $v - v'$, where the components are given by: 
\begin{equation}
    U_{ij}(v - v') = \frac{1}{|v - v'|^3} \left( \delta_{ij} - \frac{(v_i - v'_i)(v_j - v'_j)}{|v - v'|^2}  \right).
\end{equation}

The following two equations are automatically satisfied by solutions to the above system for all time: 
\begin{align}
    &\text{div } E = \frac{q}{\varepsilon_0} \int f\text{d}v,\\
    &\text{div } B = 0,
\end{align}
the latter assuming that it holds for the initial conditions.

The Vlasov-Maxwell system has a well-known Hamiltonian formulation \citep{Morrison1980, Marsden1982, Morrison:1982}, and so its coupling to dissipation in the form of the Landau operator can be formulated via the metriplectic approach. 

\subsection{The Landau Operator}

In this work, we consider purely the dissipative dynamics described by the collision operator, that is
\begin{equation}\label{eq:f_eom}
    \partial_t f = C[f] ,
\end{equation}
omitting the Hamiltonian part of the system. 
In the metriplectic framework, the collision operator $C[f]$ can be described as
\begin{equation}\label{eq:operator_bracket}
    C[f] = - (f, \mathcal S) ,
\end{equation}
where the entropy functional $\mathcal S$ is given by:
\begin{equation}\label{eq:entropy}
        \mathcal S = - \int f \log f \dx v .
\end{equation}
The metric bracket for the Landau collision operator is given by 
\begin{equation}\label{eq:landau_bracket}
(\mathcal{A},\mathcal{B}) = \frac{1}{2}\int \int \left(\frac{\partial\mathcal{A}_f}{\partial v} -\frac{\partial\mathcal{A}_{f'}}{\partial v'}  \right) f(v)U(v,v')f(v')\left(\frac{\partial\mathcal{B}_f}{\partial v} -\frac{\partial\mathcal{B}_{f'}}{\partial v'}  \right) dvdv' ,
\end{equation}
with $\mathcal{A},\mathcal{B}$ functionals, and $\mathcal{A}_f$ and $\mathcal{A}_{f'}$ denoting the functional derivatives of $\mathcal{A}$ with respect to $f(v)$ and $f(v')$, respectively.
By the chosen sign conventions for the entropy $\mathcal{S}$ and the bracket, entropy is dissipated in time:
\begin{equation}
    \frac{d}{dt}\mathcal{S} = -(\mathcal{S}, \mathcal{S}) \leq 0.
\end{equation}

There are three physically important invariants in the Vlasov system that we wish to also conserve in the metric bracket -- mass, momentum, and energy: 
\begin{equation}\label{eq:conserved_quantities_VM}
    \mathcal M = m\int f \dx v, \hspace{2em} \mathcal P = m\int vf \dx v, \hspace{2em} \mathcal E = m\int |v|^2f \dx v.
\end{equation}
We note that in the case of the full Vlasov--Maxwell--Landau system, the conserved quantities shown above are slightly modified to take the electromagnetic contributions into account.

The quantities in~\eqref{eq:conserved_quantities_VM} are Casimir invariants of the metric bracket, i.e. they satisfy: 
\begin{equation}
    (\mathcal A, \mathcal M) = 0, \hspace{2em} (\mathcal A, \mathcal P) = 0, \hspace{2em} (\mathcal A, \mathcal E) = 0, 
\end{equation}
where $\mathcal A$ is any functional. 

The equilibrium state of the system can then be obtained by applying the energy-Casimir principle \eqref{eq:energy_casimir}. Here, the equilibrium state must satisfy:
\begin{equation}
    \bigg[ \frac{\delta \mathcal S}{\delta f} + \lambda_{\mathcal M}\frac{\delta \mathcal M}{\delta f} + \lambda_{\mathcal P} \cdot \frac{\delta \mathcal P}{\delta f}  + \lambda_{\mathcal E}\frac{\delta \mathcal E}{\delta f}  \bigg] \bigg\vert_{f = f_{eq}} = 0.
\end{equation}
Using Equations~\eqref{eq:entropy} and~\eqref{eq:conserved_quantities_VM}, the above condition becomes: 
\begin{equation}
    - \left( 1 + \log f_{eq} \right) + \lambda_{\mathcal M} m + \lambda_{\mathcal P} \cdot m v + \lambda_{\mathcal E} m |v|^2 = 0,
\end{equation}
and so equilibrium distribution functions are of the form:
\begin{align}
   f_{eq} = \exp \big( -(\lambda_M m +2\lambda_P\cdot m v + \lambda_E m|v|^2) \big).
\end{align}

Thus, we have that equilibrium distribution functions of the Landau operator are Maxwellians, with mean, standard deviation, and normalisation determined by initial conditions. 

\section{Semi-discretisation}
\label{sec:particle-discretisation}

We discretise the Landau collision operator by discretising its metriplectic formulation using particles. This is to say that we do not discretise the dynamical equations, but instead we discretise the entropy and metric bracket and obtain discrete equations via the metriplectic approach detailed in Section \ref{sec:metriplectic}.
This approach consists of the following main steps: (a) discretising the distribution function, (b) discretising functionals, and (c) discretising functional derivatives, all of which will be detailed in the following.

\subsection{Approximating the distribution function}

The distribution function is represented by a collection of $N$ macro-particles (Klimontovich representation):
\begin{equation}\label{eq:klimontovich_dist}
    f_p(v, t) = \sum_{\alpha=1}^N w_\alpha \delta(v - v_\alpha(t)), 
\end{equation}
where $v_\alpha(t) \in \R^d$ is the velocity of the $\alpha$-th particle, $w_\alpha$ is its weight, and $d \in \{1, 2, 3\}$ is the number of velocity dimensions being considered. 

This representation of the distribution function, however, does not allow for an immediate discretisation of the entropy functional~\eqref{eq:entropy} as the expression resulting from evaluating entropy~$\mathcal{S}$ on the particle distribution $f_p$ is not well-defined.
To overcome this, we use the approach of \cite{Jeyakumar2023} and adopt a projected distribution function as follows: 
\begin{equation}
\label{eq:projection_f_general}
    f_s(v,t) = \sum_i \varphi_i(v) f_i,
\end{equation}
where $\{\varphi_i(v)\}_{i=1}^M$ are a set of differentiable basis functions with local/compact support in $\R^d$ and $\{f_i(v_1(t),\ldots, v_N(t))\}_{i=1}^M$ are the coefficients of the projected distribution function in this basis.

To obtain the coefficients $\{f_i\}_{i=1}^M$, we relate the two representations of the distribution function by $L^2$ projection onto the basis $\{\varphi_i\}_{i=1}^M$, requiring weak equality:
\begin{align}
    \left< f_s , \varphi_j \right> = \left< f_p , \varphi_j \right> , \quad j=1,\ldots,M ,
\end{align}
where
\begin{align*}
    \left< f_s , \varphi_j \right> &= \int f_s (v,t) \varphi_j(v) dv = \sum_{i=1}^M \, \M_{ji} f_i  , \\
    \left< f_p , \varphi_j \right> &= \int f_p(v,t) \varphi_j(v) dv = \sum_{\alpha:v_\alpha\in\Omega_{\varphi_j}} w_\alpha \varphi_j(v_\alpha) ,
\end{align*}
and $\M_{ij} = \int \varphi_i(v) \varphi_j(v) \text{d}v$ are the elements of the mass matrix $\M$ of the basis. Here, the notation $\sum_{\alpha:v_\alpha \in \Omega_{\varphi_j}}$ signifies that we only sum over the particles which are in the support of the basis function $\varphi_j$.
Thus, the coefficients of the projected distribution function are computed as follows: 
\begin{equation}\label{eq:projection_coeffs}
    f_i(v_1(t),\ldots,v_N(t)) = \sum_{k=1}^M \M_{ik}^{-1} \sum_{\alpha:v_\alpha\in\Omega_{\varphi_k}} w_\alpha \varphi_k (v_{\alpha}).
\end{equation} 
As functions of the particle velocities, these coefficients can be differentiated with respect to individual particle velocities, $v_\beta$: 
\begin{align}\label{eq:dof_derivatives}
    \frac{\partial f_i}{\partial v_\beta} = \sum_{k:v_\beta\in\Omega_{\varphi_k}} \M_{ik}^{-1} w_\beta \nabla\varphi_k(v_\beta).
\end{align}
While it is not possible to evaluate the entropy~\eqref{eq:entropy} on the particle distribution function $f_p$, evaluation on the projected distribution function $f_s$ poses no issues. This second representation of the distribution function can thus be seen as a method of regularisation in a similar vein as previous works that used radial basis functions and similar methods to this end~\citep{Carrillo2020,Hirvijoki2021}.

\subsection{Discrete functional derivatives}

Following the metriplectic formulation, we will discretise the entropy functional and bracket separately, and then apply the discrete bracket to the discrete entropy to obtain the semi-discrete equations of motion. 
Discretising the metric bracket for the Landau operator~\eqref{eq:landau_bracket} requires the discretisation of the functional derivatives appearing therein.
This is facilitated by the observation that functionals of the distribution function, $\mathcal A[f]$, can be considered as functions of the particle velocities $V = \{v_\alpha\}_{\alpha=1}^N$, upon evaluation on the particle distribution $f_p$:
\begin{equation}\label{eq:discrete_functional}
\mathcal{A}[f_p] = A(V) .
\end{equation}
By considering variations of both sides of this expression, we obtain an approximation of the functional derivatives of $\mathcal{A}[f_p]$ in terms of the function $A$ as follows\footnote{We note that, generally speaking, this approach to discretising the functional derivatives requires the functional being differentiated to be linear in the distribution function $f$. While, in theory, this may seem very restrictive, in practise, this choice has no impact on the type of ``discrete functionals'' on which the resulting bracket may be evaluated. Regardless of this ansatz, the discrete bracket will be applicable to any function of the discrete degrees of freedom, no matter if it depends linearly or nonlinearly on the degrees of freedom, and independently from how it is constructed, e.g. by evaluating a continuous functional on the particle distribution, $f_p$, or in some other way.}.
Starting with the left-hand side, variation and integration by parts gives,
\begingroup
\allowdisplaybreaks
\begin{align}
\nonumber
\delta \mathcal A[f_{p}]
&= \int \frac{\delta \mathcal A}{\delta f} \, \delta f_{p} \, dv \\
\nonumber
&= - \sum \limits_{a=1}^{N_{p}} w_a \int \dfrac{\delta \mathcal A}{\delta f} \, \nabla_v \delta \big( v - v_a (t) \big) \cdot \delta v_a \, dv \\
&= \sum \limits_{a=1}^{N_{p}} w_a \, \nabla_v \frac{\delta \mathcal A}{\delta f} \bigg\vert_{v = v_a} \cdot \delta v_a .
\end{align}
\endgroup
Upon equating this expression with the variation of the right-hand side of~\eqref{eq:discrete_functional}, namely
\begin{align}
\delta A (V)
&= \sum \limits_{a=1}^{N_{p}} \frac{\partial A}{\partial v_a} \, \delta v_a ,
\end{align}
we obtain:
\begin{equation}\label{eq:scalar_invariance_derivative}
    \frac{\partial A}{\partial v_\alpha} = w_\alpha \frac{\partial}{\partial v} \left.\frac{\delta \mathcal{A}}{\delta f}\right|_{v = v_\alpha}.
\end{equation}

Evaluating the bracket of Equation~\eqref{eq:landau_bracket} on the particle distribution from Equation~\eqref{eq:klimontovich_dist}, we obtain the following bracket on functionals:
\begin{equation}
    (\mathcal{A},\mathcal{B}) = \frac{1}{2}\sum_{\alpha,\beta} w_\alpha w_\beta \int \left(\frac{\partial\mathcal{A}_f}{\partial v} -\frac{\partial\mathcal{A}_{f'}}{\partial v'}  \right) \delta(v-v_\alpha)U(v,v')\delta(v'-v_\beta)\left(\frac{\partial\mathcal{B}_f}{\partial v} -\frac{\partial\mathcal{B}_{f'}}{\partial v'}  \right) dvdv',
\end{equation}
for some functionals $\mathcal{A}, \mathcal{B}$. Using relation~\eqref{eq:scalar_invariance_derivative} to replace the functional derivatives $\mathcal{A}_f$, etc., we obtain the discrete bracket on \textit{functions}:
\begin{equation}\label{eq:landau_bracket_discrete}
    (A,B)_h = \frac{1}{2}\sum_{\alpha,\beta} w_\alpha w_\beta\left( \frac{1}{w_\alpha}\frac{\partial A}{\partial v_\alpha} - \frac{1}{w_\beta}\frac{\partial A}{\partial v_\beta}\right) U(v_\alpha,v_\beta) \left(\frac{1}{w_\alpha}\frac{\partial B}{\partial v_\alpha} - \frac{1}{w_\beta}\frac{\partial B}{\partial v_\beta}  \right), 
\end{equation}
where $A$, $B$ are functions of the particle velocities $V = \{v_\alpha\}_{\alpha=1}^N$.

\subsection{Discrete entropy}

The remaining ingredient required for constructing the discrete collision operator is the entropy. The entropy functional~\eqref{eq:entropy} is not amenable to plain evaluation on the particle distribution $f_p$ as the resulting expression is not well-defined. 
Thus, we evaluate the entropy functional~\eqref{eq:entropy} on the projected distribution function $f_s$ instead, such that the discrete entropy becomes:
\begin{align}\label{eq:disc_entropy}
    S_h &= -\int f_s\log{f_s} \, dv 
    = -\int \left( \sum_i \varphi_i(v) f_i \right) \log \left( \sum_j\varphi_j(v) f_j\right) \dx v .
\end{align}

Using this expression for entropy, we now compute its derivative with respect to the degrees of freedom (this will be required to apply the metric bracket to the entropy):
\begin{align}
    \frac{\partial S_h}{\partial v_\alpha} = -\int \left[ \sum_i \frac{\partial f_i}{\partial v_\alpha} \varphi_i(v) \log \left( \sum_j\varphi_j(v) f_j \right) + \frac{\sum_i\varphi_i(v) f_i}{\sum_j\varphi_j(v) f_j} \sum_j\frac{\partial f_j}{\partial v_\alpha} \varphi_j(v) \right] \dx v, 
\end{align}
where the derivative $\partial f_i / \partial v_\alpha$ is given by Equation \eqref{eq:dof_derivatives}. Simplifying the second term and using Equation \eqref{eq:dof_derivatives} gives us the following: 
\begin{align}
    \nonumber
    \frac{\partial S_h}{\partial v_\alpha} &= -\int \left[ \sum_i \frac{\partial f_i}{\partial v_\alpha} \varphi_i(v) \log \left( \sum_j\varphi_j(v) f_j \right) + \sum_j\frac{\partial f_j}{\partial v_\alpha} \varphi_j(v) \right]  \dx v, \\
    \nonumber
    &= -\int \left[ \sum_{i, k} \M^{-1}_{ik} w_\alpha \nabla \varphi_k(v_\alpha) \varphi_i(v) \log \left( \sum_{j,l} \varphi_j \M^{-1}_{jl} \sum_\beta w_\beta \varphi_l(v_\beta)  \right) \right. \\
    & \hspace{6em} \left. + \sum_{j,l} \M^{-1}_{jl} w_\alpha \nabla \varphi_l(v_\alpha) \varphi_j(v) \right] \dx v. \label{eq:S_der_line2}
\end{align}

We define the following for convenience: 
\begin{align}\label{eq:defn_Lk}
    \mathbb L_{k} = \sum_i \M^{-1}_{ik} \int \varphi_i(v) \left[ 1 + \log \left( \sum_j f_j \varphi_j(v) \right)  \right] \dx v.
\end{align}
Relabelling indices in \eqref{eq:S_der_line2} and using \eqref{eq:defn_Lk}, we obtain our final expression for the derivative of the discrete entropy with respect to the degrees of freedom:
\begin{equation}\label{eq:entropy_derivative}
    \frac{\partial S_h}{\partial v_\alpha} = - \sum_k w_\alpha\mathbb{L}_{k} \nabla\varphi_{k}(v_\alpha).
\end{equation}
With this, we can now compute the collision contribution for a given particle, $\dot v_\gamma$, through the metric bracket.

\subsection{Discrete collision operator}

The discrete collisional dynamics are given by
\begin{align}\label{eq:discrete-collisional-dynamics}
\dot{v}_\gamma &= - (v_\gamma,S_h)_h .
\end{align}
Upon inserting the expression of the discrete bracket~\eqref{eq:landau_bracket_discrete} and the derivative of the entropy~\eqref{eq:entropy_derivative}, we obtain:
\begin{align}
\dot{v}_\gamma
\label{eq:eom_line1}
&= \frac{1}{2}\sum_{\alpha,\beta}w_\alpha w_\beta \left(\frac{1}{w_\alpha}\delta_{\alpha\gamma} - \frac{1}{w_\beta}\delta_{\beta\gamma}\right) U(v_\alpha,v_\beta) \sum_k\mathbb{L}_k(\nabla \varphi_{k}(v_\alpha) - \nabla \varphi_{k}(v_\beta)),\\
\nonumber
&=\frac{1}{2} \left[ \sum_\beta w_\beta U(v_\gamma,v_\beta) \sum_k \mathbb{L}_{k}(\nabla \varphi_{k}(v_\gamma) - \nabla \varphi_{k}(v_\beta)) \right. \\
\label{eq:eom_line2}
&\hspace{6em}- \left.\sum_\alpha  w_\alpha U(v_\alpha,v_\gamma) \sum_k\mathbb{L}_k (\nabla \varphi_{k}(v_\alpha) - \nabla \varphi_{k}(v_\gamma)) \right], \\ 
\label{eq:eom_line3}
&= \sum_\alpha w_\alpha U(v_\gamma,v_\alpha)\sum_k\mathbb{L}_k(\nabla \varphi_{k}(v_\gamma) - \nabla \varphi_{k}(v_\alpha)). 
\end{align}
Here, we obtain Equation \eqref{eq:eom_line2} through separating the two Kronecker delta terms in \eqref{eq:eom_line1} and contracting the appropriate indices. Equation \eqref{eq:eom_line3} then comes from relabelling indices and recognising that the two sums in \eqref{eq:eom_line2} are in fact identical by symmetry of the kernel $U(\cdot, \cdot)$. Equation \eqref{eq:eom_line3} is the final semi-discretised system of equations. 

For the purpose of discretising the system in time, it will be useful to rewrite the semi-discretised system of equations as a gradient system. To obtain this form, we rearrange Equation \eqref{eq:eom_line1} as follows with the use of \eqref{eq:entropy_derivative}: 
\begin{equation}\label{eq:landau_sd_gradient_form}
    \dot{v}_\gamma = \mathbb{G}_{\gamma \sigma} \frac{\partial S_h}{\partial v_\sigma} = -\sum_\beta\left( \frac{w_\beta}{w_\gamma}\delta_{\gamma \sigma} - \delta_{\beta \sigma} \right) U(v_\gamma, v_\beta) \frac{\partial S_h}{\partial v_\sigma}.
\end{equation}
Here, the operator $\mathbb G$ can be expressed as:
\begin{equation}\label{eq:G_semi-disc}
    \mathbb G(v_1, \dots, v_n) = \begin{bmatrix}
        -\sum\limits_{\beta \neq 1} U(v_1, v_\beta) \frac{w_\beta}{w_1} && U(v_1, v_2) && U(v_1, v_3) && \dots \\
        U(v_2, v_1) && -\sum\limits_{\beta \neq 2} U(v_2, v_\beta) \frac{w_\beta}{w_2} && U(v_2, v_3) && \dots \\
        \vdots && \hspace{7em}\ddots && && 
    \end{bmatrix} .
\end{equation}
The matrix $\mathbb G$ is a $Nd \times Nd$ block matrix, where $d$ is the number of velocity-space dimensions and $N$ the number of particles, and each $d \times d$ block is given by 
\begin{align}
\mathbb G_{\gamma \sigma} = \begin{cases}
 -\sum\limits_{\beta \neq \gamma} U(v_\gamma, v_\beta) \frac{w_\beta}{w_\gamma} \qquad & \text{if} \; \gamma = \sigma , \\
 U(v_\gamma, v_\sigma) & \text{else} .
\end{cases}
\end{align}

We observe by the symmetry of the kernel $U(\cdot, \cdot)$ that the operator $\mathbb G$ is also symmetric. It is also negative semi-definite. To see this, take a vector $p \in \R^{dN}$ and interpret it as structured into $N$ blocks $p_\gamma$ of size $d$, in analogy to the structure of $\G$, so that each block $p_\gamma \in \R^d$. Multiplying the operator $\G$ from both sides with $p$, we have: 
\begin{align}
    p^T_\gamma \G_{\gamma \sigma}p_\sigma 
    &= -\sum_{\gamma, \sigma = 1}^N \sum_{\beta=1}^N p^T_\gamma\left( \frac{w_\beta}{w_\gamma}\delta_{\gamma \sigma} - \delta_{\beta \sigma} \right) U(v_\gamma, v_\beta) p_\sigma, \nonumber  \\
    &= -\sum_{\sigma=1}^N \sum_{\beta=1}^N \frac{w_\beta}{w_\sigma} p^T_\sigma U(v_\sigma, v_\beta) p_\sigma + \sum_{\gamma=1}^N \sum_{\beta=1}^N p^T_\gamma U(v_\gamma, v_\beta)p_\beta, \nonumber \\
    &= -\sum_{\sigma=1}^N \sum_{\beta=1}^N \left(\frac{w_\beta}{w_\sigma} p^T_\sigma U(v_\sigma, v_\beta) p_\sigma -  p^T_\sigma U(v_\sigma, v_\beta) p_\beta \right).
    \label{eq:semi-def_G_expansion}
\end{align}
Here, we have expanded out the two terms, contracted the appropriate indices with the delta functions, and relabelled the resulting indices in the second sum. Factoring out the left-multiplying terms, we obtain:
\begin{align}\label{eq:semi-def_G_expansion_2}
    p^T_\gamma \G_{\gamma \sigma}p_\sigma  &= -\sum_{\sigma=1}^N \sum_{\beta=1}^N \frac{p^T_\sigma}{w_\sigma} U(v_\sigma,v_\beta)\left(w_\beta p_\sigma - w_\sigma p_\beta \right),\\
    &= -\frac{1}{2}\sum_{\sigma=1}^N \sum_{\beta=1}^N \left(w_\beta p_\sigma-w_\sigma p_\beta \right)^T \frac{U(v_\sigma,v_\beta)}{w_\sigma w_\beta}\left(w_\beta p_\sigma - w_\sigma p_\beta \right)\leq 0.\label{eq:semi-def_G_expansion_3}
\end{align}
The final equation is obtained by realising that swapping indices in Equation~\eqref{eq:semi-def_G_expansion_2} results in the same term up to a minus sign, and further re-arranging results in the form shown above. As the matrix $U$ is a projector and therefore positive semi-definite, and is being multiplied from the left and right by the same vector, we have that every term in the sum is non-positive. Thus the entire sum is non-positive and the operator $\G$ is negative semi-definite. 

\subsection{Conservation of momentum and energy}

We now demonstrate that the semi-discrete equations of motion preserve total momentum and energy. We note that mass is trivially preserved by this scheme, due to the particle weights being held constant. We start by looking at the evolution of the total particle momentum, $P = \sum_\gamma w_\gamma v_\gamma$:
\begin{align}
    \frac{\dx{}}{\dx{t}} \sum_\gamma w_\gamma v_\gamma &= \sum_\gamma w_\gamma \dot v_\gamma
    = \sum_\gamma w_\gamma \mathbb G_{\gamma \sigma} \frac{\partial S_h}{\partial v_\sigma}, 
\end{align}
which gives us that $\dx P/\dx t = 0$, as the product $w_\gamma \mathbb G_{\gamma \sigma}$ is equal to a vector which is zero element-wise. Thus, the semi-discretisation preserves total particle momentum over time. 

To demonstrate energy conservation, it will be useful to consider the action of the operator $U(\cdot, \cdot)$. We note that $U(v, v')$ is an orthogonal projector onto the subspace orthogonal to the vector $v - v'$. It follows that $U(v, v')(v - v') = U(v', v)(v - v') = \Vec 0$.  We now follow the same process as before to check conservation, where the energy density is given by $E = \sum_\gamma w_\gamma v_\gamma^2 /2$:
\begin{align}
     \frac{\dx{E}}{\dx{t}} &= \sum_\gamma w_\gamma v_\gamma \dot v_\gamma
     = \sum_\gamma w_\gamma v_\gamma \mathbb G_{\gamma \sigma} \frac{\partial S_h}{\partial v_\sigma}.
\end{align}

Without loss of generality, we choose $\sigma = 1$. If we consider the contraction $\sum_\gamma w_\gamma v_\gamma \mathbb G_{\gamma 1}$, we obtain:
\begin{align}
    \sum_\gamma w_\gamma v_\gamma \mathbb G_{\gamma 1}
    \nonumber
    &= -\sum_{\beta \neq 1} U(v_1, v_\beta)w_\beta v_1 + U(v_1, v_2)w_2 v_2 \\
    \nonumber
    &\hspace{6em} + U(v_1, v_3)w_3 v_3 + \dots + U(v_1, v_n)w_n v_n \\
    \nonumber
    &= \sum_{\beta \neq 1} w_\beta U(v_1, v_\beta)(v_\beta - v_1) \\
    &=0,
\end{align}
where the last line follows by the property of $U(\cdot, \cdot)$ being an orthogonal projector. As this argument holds for all $\sigma$, it follows that $\dx E/ \dx t = 0$ and the semi-discretisation preserves the energy.

\subsection{H-Theorem}

The semi-discrete collisional dynamics~\eqref{eq:discrete-collisional-dynamics} satisfies a H-theorem, such that entropy is monotonically dissipated. 
This can be seen by considering the time evolution of the discretised entropy:
\begin{align}
    \nonumber
    \frac{d}{dt}S_h &= -(S_h, S_h)_h \\
    &=  -\frac{1}{2}\sum_{\alpha,\beta} w_\alpha w_\beta\left( \frac{1}{w_\alpha}\frac{\partial S_h}{\partial v_\alpha} - \frac{1}{w_\beta}\frac{\partial S_h}{\partial v_\beta}\right)^T U(v_\alpha,v_\beta) \left(\frac{1}{w_\alpha}\frac{\partial S_h}{\partial v_\alpha} - \frac{1}{w_\beta}\frac{\partial S_h}{\partial v_\beta}  \right),
\end{align}
where $(\cdot, \cdot)_h$ is the semi-discrete metric bracket given in Equation~\eqref{eq:landau_bracket_discrete}. 

As the matrix $U$ is positive semi-definite, and the terms multiplying it in the above equation are identical for a given $\alpha$ and $\beta$, and $w_\alpha, w_\beta \geq 0$, each term in the sum is non-negative. As such, we have that $dS_h/dt  \leq 0$ and so the discretised entropy is monotonically dissipated. 

We can also compute the discrete equilibrium state by using the energy-Casimir principle, as before. The energy-Casimir principle for the discrete system reads: 
\begin{equation}
    \delta S_h(f_{s,eq}) + \delta \sum_i \lambda_i C_i (f_{s,eq}) = 0,
\end{equation}
where $S_h(f_{s,eq})$ is the discrete entropy \eqref{eq:disc_entropy}, evaluated at the discrete equilibrium distribution $f_{s,eq}$, that is the spline representation of the particles with equilibrated velocities $v_{\alpha,eq}$. The $C_i$'s are the discrete Casimir invariants, i.e. the discrete mass, momentum, and energy. Thus, the above equation becomes: 
\begin{equation}
    \left[ \delta S_h + \lambda_{M_h} \delta \left( \sum_\alpha w_\alpha \right) + \lambda_{P_h} \delta \left( \sum_\alpha w_\alpha v_\alpha \right) + \lambda_{E_h} \delta \left( \sum_\alpha w_\alpha v_\alpha^2 \right) \right]_{v_\alpha = v_{\alpha,eq}} = 0,
\end{equation}
where $\lambda_{M_h}$, $\lambda_{P_h}$, and $\lambda_{E_h}$ are the corresponding Lagrange multipliers for the discrete quantities. Computing the variations in this equation, we have:
\begin{equation}
   \left. \sum_\alpha \left[ \frac{\partial S_h}{\partial v_\alpha} + \lambda_{P_h}w_\alpha + \lambda_{E_h}w_\alpha v_\alpha \right]\delta v_\alpha \right|_{v_\alpha = v_{\alpha,eq}} = 0,
\end{equation}
where the quantity $\partial S_h/\partial v_\alpha$ is given by Equation~\eqref{eq:entropy_derivative}, and the term corresponding to the mass does not appear as it does not depend on the particle velocities, thus having zero variation. Thus, we find that an equilibrium distribution's degrees of freedom, i.e. the particle velocities $\{v_{\alpha, eq}\}$ must satisfy:
\begin{equation}
    \left[\frac{\partial S_h}{\partial v_\alpha} + \lambda_{P_h}w_\alpha + \lambda_{E_h}w_\alpha v_\alpha\right]_{v_\alpha = v_{\alpha,eq}} = 0,
\end{equation}
for the appropriate Lagrange multipliers based on the chosen initial conditions.

\section{Time discretisation}
\label{sec:time-discretisation}

In order to discretise the system of ODEs in time, we use the discrete gradient method \citep{Quispel1996, McLachlan1999}. It is an integral preserving method, initially developed for Hamiltonian systems and others with first integrals, which can also be adopted to dissipative systems as shown below. 

To illustrate, we begin with a system of ODEs in gradient form, $\dot z = G(z) \nabla S(z)$, where $z \in \R^n$. Here, $G(z)$ can be a skew-symmetric matrix in the case of a Hamiltonian system, or a negative semi-definite symmetric matrix in the case of a dissipative one. The system of ODEs is then approximately solved by the following equation:
\begin{equation}\label{eq:disc_grad_approx}
    \frac{z_{n+1} - z_n}{\Delta t} = \tilde G(z_n, z_{n+1}) \bar \nabla S (z_n, z_{n+1}),
\end{equation}
where $\bar \nabla S (z_n, z_{n+1})$ is known as the discrete gradient, and $\tilde G(z_n, z_{n+1})$ is a matrix such that $\tilde G(z_n, z_{n+1}) \to G(z_n)$ in the limit $z_{n+1} \to z_n$ as $\Delta t \to 0$. The discrete gradient, $\bar \nabla S (z_n, z_{n+1})$, satisfies the following two properties: 
\begin{subequations}\label{eq:disc_grad_props}
\begin{align}
    (z_{n+1} - z_n)\cdot \bar \nabla S (z_n, z_{n+1}) &= S(z_{n+1}) - S(z_n), \\
    \bar \nabla S (z_n, z_n) &= \nabla S (z_n).
\end{align}
\end{subequations}
One such discrete gradient is the one by Gonzalez \citep{Gonzalez1996}, also known as the midpoint discrete gradient: 
\begin{multline}
    \bar \nabla S (z_n, z_{n+1})
    = \nabla S (z_{n + 1/2}) \\
    + (z_{n+1} - z_n)\frac{S(z_{n+1}) - S(z_n) - (z_{n+1} - z_n)\cdot \nabla S (z_{n + 1/2})}{|z_{n+1} - z_n|^2},
\end{multline}
where $z_{n + 1/2} = (z_n + z_{n+1})/2$ is the midpoint. 

In the case of the Landau operator, Equation~\eqref{eq:landau_sd_gradient_form} is the gradient form of the semi-discrete system of ODE's. Applying the discrete gradient approximation of Equation~\eqref{eq:disc_grad_approx}, we have: 
\begin{align}\label{eq:disc_grad_update_rule}
    \frac{v_\gamma^{n+1} - v_\gamma^n}{\Delta t} = \tilde{\mathbb G}_{\gamma \sigma}(v_n, v_{n+1}) \bar \nabla_\sigma S_h (v_n, v_{n+1}),
\end{align}
where $\bar \nabla_\sigma S_h (v_n, v_{n+1})$ represents the discrete gradient of $\partial S_h/\partial v_\sigma$, satisfying the properties in \eqref{eq:disc_grad_props}, and $v_n$ represents a vector of all particle velocities at time $n$. Here, we take the midpoint approximation for the matrix $\tilde{\mathbb G}_{\gamma \sigma}$, i.e.:
\begin{equation}\label{eq:midpoint_G}
    \tilde{\mathbb G}_{\gamma \sigma}(v_n, v_{n+1}) = \mathbb G_{\gamma \sigma }\left( \frac{v_n + v_{n+1}}{2} \right).
\end{equation}
In the treatment of the full Vlasov--Maxwell-Landau system, this approach for time discretisation of the dissipative part can immediately be adopted for the conservative part as well, possibly in conjunction with some kind of operator splitting.

\subsection{Conservation of momentum and energy}
Under this scheme, we can show that momentum and energy are conserved in time as well. We first consider the total particle momentum: 
\begin{align}
    \sum_\gamma w_\gamma v_\gamma^{n+1} - \sum_\gamma w_\gamma v_\gamma^{n}
    \nonumber
    &= \Delta t \sum_\gamma w_\gamma \tilde{\mathbb G}_{\gamma \sigma}(v_n, v_{n+1}) \bar \nabla_\sigma S_h (v_n, v_{n+1}) \\
    &= \Delta t \sum_\gamma w_\gamma \mathbb G_{\gamma \sigma}\left( \frac{v_n + v_{n+1}}{2} \right) \bar \nabla_\sigma S_h (v_n, v_{n+1}), 
\end{align}
where we have used Equation~\eqref{eq:midpoint_G} to obtain the second line. By the structure of the matrix $\mathbb G_{\gamma \sigma}$ as shown in Equation~\eqref{eq:G_semi-disc}, we have that the sum $\sum_\gamma w_\gamma \mathbb G_{\gamma \sigma}(\cdot) = 0$, and so:
\begin{equation}
   \Delta P =  \sum_\gamma w_\gamma v_\gamma^{n+1} - \sum_\gamma w_\gamma v_\gamma^{n} = 0.
\end{equation}
Thus, total particle momentum is conserved exactly in time. 

Now, we consider the total particle energy. We have:
\begin{align}
    \frac{1}{2}\sum_\gamma w_\gamma \left(v_\gamma^{n+1}\right)^2
    \nonumber
    &=\frac{1}{2} \sum_\gamma w_\gamma \left[ v_\gamma^n + \Delta t \tilde{\mathbb G}_{\gamma \sigma}(v_n, v_{n+1}) \bar \nabla_\sigma S_h (v_n, v_{n+1})\right]^2, \\
    &= \frac{1}{2}\sum_\gamma w_\gamma \left(v_\gamma^{n}\right)^2 + \Delta t \sum_\gamma w_\gamma v_\gamma^n \tilde{\mathbb G}_{\gamma \sigma}(v_n, v_{n+1}) \bar \nabla_\sigma S_h (v_n, v_{n+1}) \nonumber \\
    & \hspace{3em} + \frac{1}{2}\Delta t^2 \sum_\gamma w_\gamma \left( \tilde{\mathbb G}_{\gamma \sigma}(v_n, v_{n+1}) \bar \nabla_\sigma S_h (v_n, v_{n+1}) \right)^2. \label{eq:disc_energy_proof_1}
\end{align}
Rewriting the second term of the above equation using the discrete gradient update rule rearranged as $v_\gamma^n = v_\gamma^{n+1} -\Delta t \tilde{\mathbb G}_{\gamma \sigma}(v_n, v_{n+1}) \bar \nabla_\sigma S_h (v_n, v_{n+1}) $, we have the following useful expression
\begingroup
\allowdisplaybreaks
\begin{align}\label{eq:disc_energy_proof_2}
    &\Delta t \sum_\gamma w_\gamma v_\gamma^n \tilde{\mathbb G}_{\gamma \sigma}(v_n, v_{n+1}) \bar \nabla_\sigma S_h (v_n, v_{n+1}) \nonumber \\
    &\hspace{3em}
    = \Delta t \sum_\gamma w_\gamma \left( v_\gamma^{n+1} - \Delta t \tilde{\mathbb G}_{\gamma \beta}(v_n, v_{n+1}) \bar \nabla_\beta S_h (v_n, v_{n+1}) \right)\tilde{\mathbb G}_{\gamma \sigma}(v_n, v_{n+1})\bar \nabla_\sigma S_h (v_n, v_{n+1}), \nonumber \\
    &\hspace{3em}
    = \Delta t \sum_\gamma w_\gamma v_\gamma^{n+1}\tilde{\mathbb G}_{\gamma \sigma}(v_n, v_{n+1})\bar \nabla_\sigma S_h (v_n, v_{n+1}) \nonumber \\
    &\hspace{6em}
    - \Delta t^2 \sum_\gamma w_\gamma \left( \tilde{\mathbb G}_{\gamma \sigma}(v_n, v_{n+1})\bar \nabla_\sigma S_h (v_n, v_{n+1}) \right)^2. 
\end{align}
\endgroup
Now, we rewrite Equation~\eqref{eq:disc_energy_proof_1} by splitting up the second term into two halves. Using Equation~\eqref{eq:disc_energy_proof_2} to replace one of those terms, we have:
\begingroup
\allowdisplaybreaks
\begin{align}\label{eq:disc_energy_proof_3}
    \Delta E
    &= \frac{1}{2}\sum_\gamma w_\gamma \left(v_\gamma^{n+1}\right)^2 - \frac{1}{2}\sum_\gamma w_\gamma \left(v_\gamma^{n}\right)^2 \nonumber \\
    &= \frac{1}{2}\Delta t \sum_\gamma w_\gamma v_\gamma^n \tilde{\mathbb G}_{\gamma \sigma}(v_n, v_{n+1}) \bar \nabla_\sigma S_h (v_n, v_{n+1}) \nonumber \\
    &\hspace{3em}
    + \frac{1}{2}\Delta t \sum_\gamma w_\gamma v_\gamma^{n+1}\tilde{\mathbb G}_{\gamma \sigma}(v_n, v_{n+1})\bar \nabla_\sigma S_h (v_n, v_{n+1}) \nonumber \\
    &\hspace{3em}
    - \frac{1}{2}\Delta t^2 \sum_\gamma w_\gamma \left( \tilde{\mathbb G}_{\gamma \sigma}(v_n, v_{n+1})\bar \nabla_\sigma S_h (v_n, v_{n+1}) \right)^2 \nonumber \\
    &\hspace{3em}
    + \frac{1}{2}\Delta t^2 \sum_\gamma w_\gamma \left( \tilde{\mathbb G}_{\gamma \sigma}(v_n, v_{n+1}) \bar \nabla_\sigma S_h (v_n, v_{n+1}) \right)^2, 
\end{align}
\endgroup
where the last two terms in the above equation cancel exactly. Factorising the first two terms and then using the definition of $\tilde{\mathbb G}_{\gamma \sigma}(\cdot, \cdot)$, we have: 
\begin{align}
     \Delta E
     \nonumber
     &= \Delta t \sum_\gamma w_\gamma \left(\frac{v_\gamma^n + v_\gamma^{n+1}}{2} \right)\tilde{\mathbb G}_{\gamma \sigma}(v_n, v_{n+1}) \bar \nabla_\sigma S_h (v_n, v_{n+1}) , \\
    &= \Delta t \sum_\gamma w_\gamma \left(\frac{v_\gamma^n + v_\gamma^{n+1}}{2} \right) \mathbb G_{\gamma \sigma}\left( \frac{v_n + v_{n+1}}{2} \right)\bar \nabla_\sigma S_h (v_n, v_{n+1}).
\end{align} 
 By the arguments given in the semi-discrete energy conservation proof, we have that $\sum_\gamma w_\gamma v_\gamma \mathbb G_{\gamma \sigma} = 0$, where $v_\gamma$ and $\mathbb G_{\gamma \sigma}$ are evaluated at the same point in time. As such, we have that $\sum_\gamma w_\gamma [(v_\gamma^n + v_\gamma^{n+1})/2] \mathbb G_{\gamma \sigma}[(v_\gamma^n + v_\gamma^{n+1})/2]  = 0$, and so:
\begin{equation}
    \Delta E = \frac{1}{2}\sum_\gamma w_\gamma \left(v_\gamma^{n+1}\right)^2 - \frac{1}{2}\sum_\gamma w_\gamma \left(v_\gamma^{n}\right)^2 = 0. 
\end{equation}
Thus, the total particle energy is also conserved exactly in time. 

\subsection{H-theorem}
We can demonstrate that entropy is also monotonically dissipated in time. From Equation~\eqref{eq:disc_grad_props}, we have that: 
\begin{equation}
    S(v_{n+1}) - S(v_n) = (v_\gamma^{n+1} - v_\gamma^n) \cdot \bar \nabla_\gamma S_h (v_n, v_{n+1}).
\end{equation}
Using the discrete gradient update rule \eqref{eq:disc_grad_update_rule}, this becomes:
\begin{equation}
    S(v_{n+1}) - S(v_n) = \Delta t \bar \nabla_\sigma S_h (v_n, v_{n+1}) \tilde \G_{\sigma \gamma} \bar \nabla_\gamma S_h (v_n, v_{n+1}).
\end{equation}
As the matrix $\tilde \G$ is simply $\G$ evaluated at the midpoint (c.f. Equation~\eqref{eq:midpoint_G}), it is also negative semi-definite. Since the vectors multiplying $\tilde \G$ from the left and right are identical, we have:
\begin{equation}
    S(v_{n+1}) - S(v_n) \leq 0,
\end{equation}
and so entropy is monotonically dissipated in time. Thus, the H-theorem is maintained at the discrete level.  

\section{Conclusion}

In this work, we have described a structure-preserving particle method for computing the collisional dynamics described by the Landau operator, using the metriplectic formulation. This method is based on a projection based approach, which was introduced in \cite{Jeyakumar2023}. The method's structure-preserving properties have been demonstrated analytically in the semi-discrete setting (only spatial discretisation), as well as in the fully discrete setting (including discretisation in time using discrete gradients). The H-theorem has also been shown to hold in both settings. This method can be coupled to an appropriate particle method for the Vlasov--Poisson or Vlasov--Maxwell equations, and the coupling methodology and benefits will be reported in future papers.